\documentclass[journal,article,submit,pdftex,moreauthors]{Definitions/mdpi}

\usepackage[normalem]{ulem}
\usetikzlibrary{positioning}

\firstpage{1} 
\pubvolume{1}
\issuenum{1}
\articlenumber{0}
\pubyear{2024}
\copyrightyear{2024}
\datereceived{ } 
\daterevised{ } 
\dateaccepted{ } 
\datepublished{ } 
\hreflink{} 

\PassOptionsToPackage{dvipsnames}{xcolor} 
\usepackage{xcolor}

\Title{X-ray variability of VHE detected FSRQs: A comparative study}

\TitleCitation{X-ray variability of VHE detected FSRQs: A comparative study}

\Author{Lekshmi Dathan \orcidA{} $^{1}$and Pankaj Kushwaha \orcidB{} $^{1}$ }

\AuthorNames{Lekshmi Dathan and Pankaj Kushwaha}

\AuthorCitation{Dathan, L.; Kushwaha, P.}

\address{%
$^{1}$ \quad Department of Physical Sciences, Indian Institute of Science Education and Research (IISER) Mohali, Knowledge City, Sector 81, SAS Nagar 140306, India \\
}

\corres{Correspondence: dathanlekshmi@gmail.com (L.D.) \& pankaj.kushwaha@iisermohali.ac.in (P.K.); 
}

\abstract{Flat Spectrum Radio Quasars (FSRQs) are weak sources of very high energy (VHE; E$>$100 GeV) emission, despite exhibiting strong MeV-GeV emissions that dominate their radiative output. To date, only ten FSRQs have been detected at VHEs, primarily during bright optical phases. In this study, we perform a detailed and systematic, temporal, and spectral analysis of the nine VHE-detected FSRQs, using the {\it Swift} X-ray Telescope (XRT) data. Our findings show no correlation between VHE activity and the X-ray flux or spectral state of the sources. However, investigation of spectral properties with X-ray brightness shows anti-correlation between flux and spectral index. The X-ray, generally with a different spectral shape lies at the farther end of the optical-UV synchrotron spectrum which typically shows a declining power-law spectrum, and thus, the X-ray spectrum is generally explained by Synchrotron Self-Compton (SSC) process. However, if optical-UV synchrotron emission extends into the X-ray band, it can soften the X-ray spectrum. While most sources in our sample exhibit rising X-ray SEDs, indicative of non-synchrotron origins or minimal synchrotron contributions, many display softer or flat X-ray spectra, mainly during low X-ray flux states (e.g., 4C +21.35, 3C 279, TON 0599, PKS 1441+25, and PKS 0346-27)  suggesting potential synchrotron contributions. These synchrotron continuations influence the gamma-ray spectrum, implying extension into the VHE range for inverse Compton (IC) scattering in the Thomson scattering limit. If the extended component corresponds to an underlying low-level emission, these FSRQs could represent potential candidates for persistent VHE activity.}


\begin{document}
\nolinenumbers

\section{Introduction}

Blazars are a sub-type of radio-loud Active Galactic Nuclei (AGNs) which have relativistic jets oriented at a very small angle ($\leq 15 ^0$) with respect to the line of sight \textbf{\citep{1995PASP..107..803U, 2013MNRAS.431.1914G, 1978PhyS...17..265B}}. They have two sub-categories - BL Lacartae objects (BL Lacs) and Flat Spectrum Radio Quasars (FSRQs) based on the presence or absence of strong emission line features in their optical-UV spectra \citep{1995PASP..107..803U} which are characterized by the rest-frame equivalent width (EW). BL Lacs have very weak (EW $<$ 5\AA) or no emission lines, whereas FSRQs have prominent (EW $\geq$ 5\AA) emission lines in their spectra \citep{1991ApJS...76..813S}. Blazars are very bright with highly variable emission in all wavebands from radio up to GeV/TeV $\gamma$ rays \citep{1996ASPC..110.....M, 1997ARA&A..35..445U, 2007ApJ...664L..71A, 2020MNRAS.496.1430P}.
Many of them have been detected at Very High Energy (VHE, $E>100$ GeV), and these are termed Very High Energy Blazars or VHE Blazars \citep{2011ICRC....8...47B}. \\

The standard blazar emission paradigm consists of a supermassive black hole that has relativistic jets of magnetized plasma ejected in our line of sight. The broadband emission from the jets we know now to be non-thermal in origin and the broad-band Spectral Energy Distribution (SED) has a characteristic double-humped structure \citep{1999APh....11..159U}. The first peak, or the low-energy hump peaks between infrared (IR) to X-ray energies and is widely accepted to be caused by the synchrotron emission from the relativistic electrons in the jet \citep{1995PASP..107..803U}. The second or the high-energy hump peaks at the MeV-GeV $\gamma$-ray energies \citep{2017SSRv..207....5R}. The cause of these high energy $\gamma$-ray emissions is still under debate, with both leptonic and hadronic processes involving relativistic particles have been suggested for $\gamma$-ray emission \citep{2000ApJ...544L..23T}. However, a fully hadronic scenario has been found energetically unfavourable within the basic accretion paradigm requiring super Eddington power \citep{2015MNRAS.450L..21Z} and also from modelling the broadband SED of the first candidate neutrino blazar TXS0506+056 \citep{2019NatAs...3...88G}. \\

In the leptonic scenario, the low-energy-hump i.e., the radio-to-optical emission, is due to the synchrotron emission from relativistic electrons within the jet, and the second or the high-energy hump that peaks at the $\gamma$-ray energy is attributed to IC scattering \citep{1997ApJ...484..108S}. The most effective seed photons for the IC scattering process are either provided by the synchrotron emission which causes Synchrotron Self-Compton (SSC) \citep{1992ApJ...397L...5M, 1994IAUS..159..155M, 1998MNRAS.301..451G} or from other parts of the AGN, such as the accretion disk \citep{1993ApJ...416..458D}, BLR region \citep{1994ApJ...421..153S}, molecular torus \citep{1995ARA&A..33..163W} or also the CMB \citep{2008ApJ...679L...9B}, which are broadly termed as External Comptonization (EC). \\


Blazars are also classified based on the frequency at which the low-energy hump or the synchrotron emission peaks ($\nu_{peak}$). Depending upon the peak frequency, it is divided into low-synchrotron-peaked (LSP, $\nu_{peak}<10^{14} Hz$), intermediate-synchrotron-peaked (ISP, $<10^{14}\nu_{peak}<10^{15} Hz$), and high-synchrotron-peaked (HSP, $\nu_{peak}>10^{15"} Hz$) blazars
\citep{2010ApJ...716...30A,1998MNRAS.299..433F}. FSRQs come under the category of LSPs and their radiative output is dominated by the MeV-GeV $\gamma$ rays. However, it is interesting to note that the extra-galactic catalogue of VHE emissions is largely dominated by high-frequency peaked BL Lacs while FSRQs despite being strong MeV-GeV gamma-ray emitters are hardly detected at VHEs.
Out of a total of 651 FSRQ detected in the 4th Catalogue of the Fermi gamma-ray observatory, only 10 of them have been found to emit Very High Energies (Refer Table ~\ref{table:FSRQ} \citep{2020ApJS..247...33A}). \\

Several factors contribute to the low number of FSRQs detected at VHE. FSRQs are LSPs and hence the peak of the gamma-ray emission in their SED tends to shift to lower energies compared to BL Lac objects \citep{2016Galax...4...36G}. Enhanced internal absorption within the Broad Line Region (BLR) (eg. \citep{2006ApJ...653.1089L}, \citep{2018MNRAS.477.4749C}), due to pair production is another factor. It is also important that VHE gamma-ray emissions from FSRQs to date have happened at either brief flaring events or extended high states \citep{2021A&A...647A.163M}. PKS 1510-089 is an exception from which persistent VHE gamma-ray emission is observed during low High Energy (HE, E > 10 GeV) state \citep{2023ApJ...953...47Y}. Since there are currently relatively few known VHE FSRQs, it is crucial to study those objects to identify any similarities or differences in their emissions and to check if the same processes are responsible. This study focuses particularly on the X-ray emissions. \\

In LSPs, to which FSRQs belong, X-ray emission is due to IC scattering of the synchrotron photons i.e. (SSC), but the lower-energy part of the X-ray spectrum ($\lesssim 2$ keV) can have a substantial contribution from the high energy tail of the synchrotron emission; this is when the low-energy X-ray emission falls at the extrapolation of optical-UV synchrotron flux \citep{2016MNRAS.458...56W}. Significant contribution at low energy part of the X-ray has been reported in many LSP BL Lacs (e.g. \citep{2022MNRAS.509.2696S,2024arXiv241005783K}) while hardening of the low-energy X-ray has been seen in some VHE FSRQs, which could be due to synchrotron \citep[e.g.][]{2015ApJ...815L..23A,2016MNRAS.458...56W} or could be a new HSP-like emission component \citep[e.g.][]{2021A&A...647A.163M}. The X-ray emission due to IC by CMB photons (EC-CMBR) is also proposed \citep{2000ApJ...544L..23T,2002ApJ...571..206S,2020Galax...8...71P}, but it was not adequate to explain the large-scale gamma-ray emission in quasar 3C 273 \citep{2014ApJ...780L..27M}. Even in the case of the SSC scenario, it requires an unusually low magnetic field \citep{2009ApJ...703.1168B} in order to explain the VHE emission. Therefore the second hump is plausibly due to EC. But then, if we interpret that X-ray emission is also due to the EC, this would pose a problem, as it would require a magnetic field much lower than its equipartition value. Hence, considering X-ray emission to be due to SSC and gamma-ray emission to be due to EC, in the VHE emission cases, can resolve this issue \citep{2017SSRv..207....5R}. For FSRQs, the emission in the MeV-GeV happens due to the IC of BLR photons from the BLR region and IR photons from the dusty torus. The EC due to BLR lies in the Klein-Nishina region and a sharp cut-off happens typically around 20-30 GeVs. So, IR photons are required for the EC in VHE scenarios, and VHE corresponds to the very high energy end of the particle spectrum (e.g. \citep{2023ApJ...952L..38A}). \\

In the currently favoured leptonic origin scenario, VHE emission in FSRQs could result either due to a new HBL-like emission component \citep[e.g.][]{2016A&A...595A..98A}, or in one zone scenario, shift in SED peak due to the break energy of the particle distribution used for SED modelling or beaming \citep[e.g.][]{2021A&A...647A.163M}, or a much higher synchrotron high-energy tail leading to IC-IR to VHE energies \citep[e.g.][]{2015ApJ...815L..23A,2017MNRAS.470.2861S}. In the latter case, the resulting VHE spectrum is directly related to the optical-UV spectrum. However, such tails can affect the low-energy part of the X-rays \citep[e.g.][]{2015ApJ...815L..23A}. Thus, investigation of X-rays during VHE activity vis-a-vis non-VHE episodes holds a potential tool to explore clues/connection with VHE emission. In such cases, the lower end of the X-ray spectrum is expected to be relatively softer in general, either due to synchrotron soft-tail or an HBL-like component. It should be noted that in terms of energy output, MeV-GeV contribution is, in general, an order of magnitude or higher than that of the respective synchrotron maximum flux i.e. IR-NIR part, normally referred to as the Compton dominance (CD) in literature \citep{2013ApJ...763..134F}. Thus, contrary to one zone, even if an additional HSP-like component is present but has CD-like FSRQs, the characteristic softer X-ray spectrum may not reflect observationally  \citep[e.g.][]{2016A&A...595A..98A} but in such cases MeV-GeV is expected to hold the clue with spectrum tending to an HSP-like component providing sufficient brightness allowing spectra beyond 30-40 GeVs. \\

In this paper, we perform a systematic investigation of the spectral and temporal variability of the 9 FSRQs that have been detected at VHEs (Refer Table \ref{table:FSRQ}). They are studied in their X-ray regime using the observations made by the X-ray Telescope (XRT) of the Neils Geherel Swift Observatory. The aim is to find the similarities and differences between these sources and also the peculiarity of them at the time of VHE emission and other times. The details of the data analysis are presented in section \ref{Data Reduction}. Section \ref{AR} presents analysis and results followed by discussion in section \ref{dis}. We finally conclude and summarize our work in section \ref{con}.

\begin{table}[H] 
\caption{Basic details: RA, DEC and redshift, of the FSRQs detected at VHE}
\begin{tabularx}{\textwidth}{CCCC}
     \toprule
     \textbf{Source name} & \textbf{RA} & \textbf{Dec} & \textbf{Redshit(z)}\\ 
     \midrule
     PKS 0736+017 & 07 39 17.0 & +01 36 12 & 0.18941 \\ 
     PKS 1510-089 & 15 12 52.2 & -09 06 21.6 & 0.361\\ 
      4C +21.35 & 12 24 54.4 & +21 22 46 & 0.432 \\ 
     3C 279 & 12 56 11.1 & -05 47 22 & 0.5362 \\ 
     B2 1420+32  & 14 22 30.38 & +32 23 10.44 & 0.682 \\ 
     TON 0599 & 11 59 31.8 & +29 14 44 &  0.7247\\ 
     PKS 1441+25 & 14 43 56.9 & +25 01 44 & 0.939 \\ 
     S3 0218+35 & 02 21 05.5 & +35 56 14 & 0.954 \\ 
     PKS 0346-27 & 03 48 38 & -27 49 14 & 0.991 \\
     \bottomrule
\end{tabularx}
\label{table:FSRQ}
\end{table}

\section{XRT Data Reduction}
\label{Data Reduction}

The X-ray Telescope (XRT) is one among the three telescopes onboard the {\it Swift } observatory \citep{2004ApJ...611.1005G}. It is sensitive in the energy range of 0.3-10 keV. {\it Swift} is designed to operate autonomously and the instruments onboard like the XRT have multiple modes of operation capable of sensitively recording a wide range of brightness state of an astronomical phenomenon. It was primarily designed for gamma-ray Bursts (GRBs) observation and thus, is more inclined towards transient brightening events, making it one of the best transient multi-band observing facilities. \\

The paper utilizes the data taken by the telescope in the time period 2006 - 2022 (MJD 54000-60000) which is publicly available. The data is analyzed using the HEASOFT package version 6.30.1 along with CALDB. All pointing observations made with \begin{em}
    Swift
\end{em}-XRT in the Photon counting (PC) and Windowed Timing (WT) mode during the period is considered. \\ 

First, all the XRT event files were re-processed using the standard filtering procedure of XRTPIPELINE \footnote{http://www.swift.ac.uk/analysis/xrt/} and calibrated using the latest calibration files from the \begin{em}
    Swift
\end{em} CALDB (with update till 31/03/2022). The source photons for the analysis were extracted using a circular region of 20 pixels $(\sim 47")$ \citep{2005SPIE.5898..360M} centred at the source. Background photons were extracted using an annular region around the source, excluding the source region. Auxillary response files (ARF) were generated using \begin{em}
    xrtmkarf
\end{em} which accounts for the CCD defects and PSF corrections. The data in the energy bins of the spectrum files were re-binned using \begin{em}
    GRPPHA
\end{em} to have at least 20 counts per bin for spectral analysis with the statistics. \begin{em}
    XSPEC
\end{em} (v.12.12.1) was used to fit the spectra to a power law 
\begin{equation}\label{eq1}
    \frac{dN}{dE} = KE^{-\alpha}
\end{equation}
for all sources except 4C $+$ 21.35 where a broken power law was used.
\begin{equation}
\frac{dN}{dE} = \begin{cases} 
KE^{-\alpha_1} & \text{if } E \leq E_{\text{break}} \\ 
KE_{\text{break}}^{\alpha_2 - \alpha_1} \left(E/1 \, \text{keV}\right)^{-\alpha_2} & \text{if } E > E_{\text{break}}
\end{cases}
\end{equation}
\\
where, \\
$\alpha$ - photon index of the power laws \\
$\alpha_1$ - power law photon index for $E < E_{break}$ \\
$\alpha_1$ - power law photon index for $E > E_{break}$ \\
$E_{break}$ - break point for the energy in keV \\
K - photons/keV/cm$^2$/s at 1 keV.

\section{Analysis and Results}
\label{AR}

\begin{figure}[hbt!]
\begin{adjustwidth}{-\extralength}{0cm}
\centering\includegraphics[width=15.5cm]{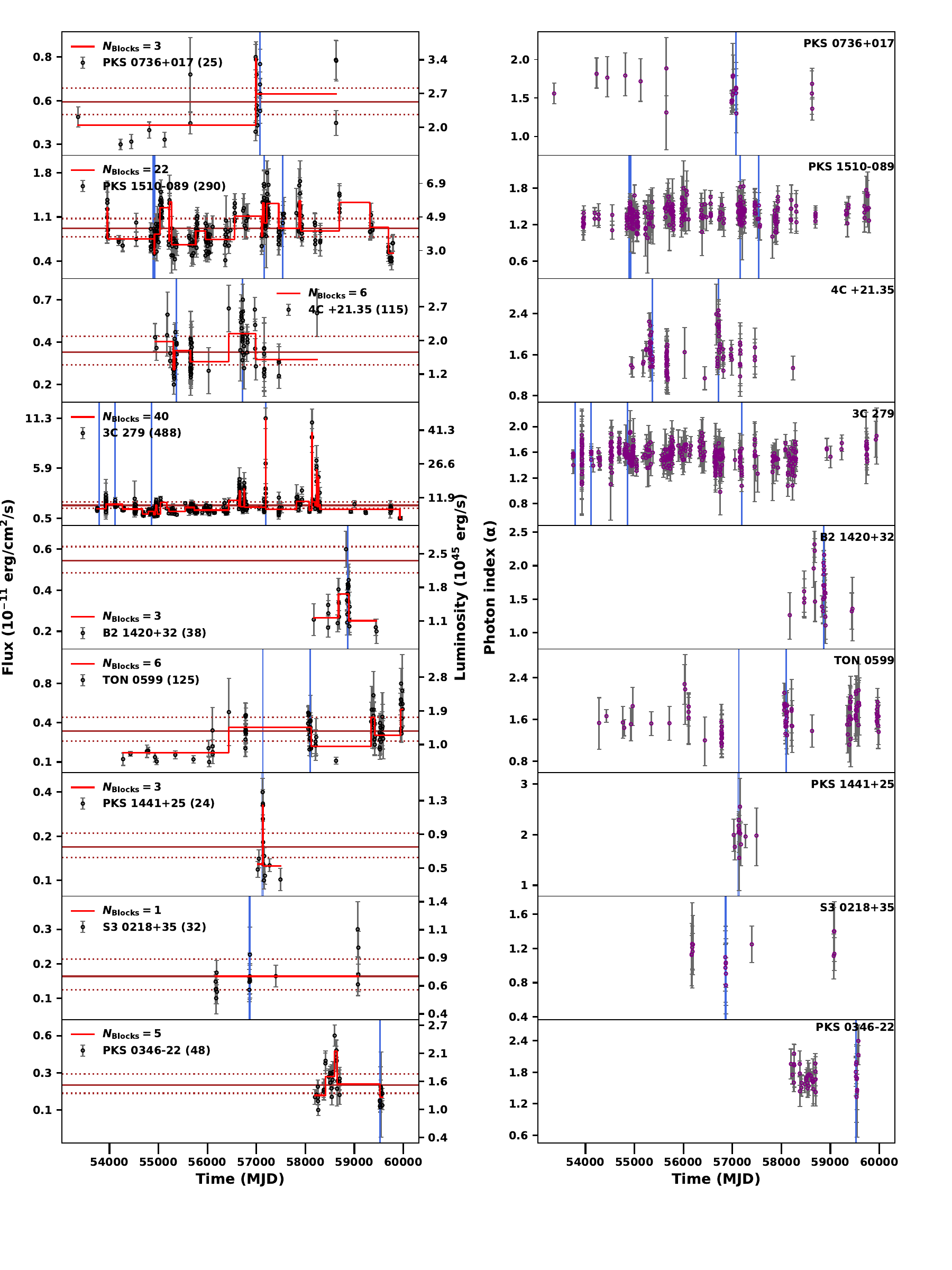}
\vspace{-1cm}
\caption{Light Curve of all the nine FSRQs in increasing order of redshift (ref. Table \ref{table:FSRQ}) from top to bottom. The left plot shows the Flux vs time (in MJD) with mean and mean error (1 $\sigma$) plotted by brown solid and brown dotted lines, respectively. The red line shows the Bayesian Blocks. The number of data points is labelled against the source name in the brackets. The right plot shows the spectral energy index vs Time (in MJD). The blue line or portion represents the VHE detection time.}
\label{fig:lc}
\end{adjustwidth}
\end{figure}

The X-ray light curve of the nine VHE FSRQs extracted following the procedures mentioned above is shown in the left panel of Figure \ref{fig:lc}. The corresponding photon spectral indices are shown in the right panel. In addition to the mean X-ray flux and the {$1\sigma$} band around it, marked by the brown solid and dashed lines, respectively, we have used Bayesian block \citep{2013ApJ...764..167S} (red solid lines) to identify flux changes. The time of VHE detections/activities reported by the ground-based VHE facilities: MAGIC, HESS, and VERITAS are marked by the blue vertical solid lines. For the duration considered in this work (2006 - 2022; MJD 54000-60000), the reported times of VHE activity are as follows: PKS 0736+017 (MJD: 57072), PKS 1510-089 (MJD: 54891-54926, 57153-57167, 57538), 4C+21.35 (MJD: 55364, 56714-56726), 3C 279 (MJD:  57184-57194, 54116, 54863, 53789), B2 1420+32 (MJD: 58868), TON 0599 (MJD: 58102, 57129-57131), PKS 1441+25 (MJD:57113, 57142), S3 0218+35 (MJD: 56861, 56864), PKS 0346-22 (MJD: 59521). \\

It is important to note here that as already mentioned Swift observation is more inclined towards transient events, hence the data collected will be skewed to include observations of sudden brightening. 
This is apparent from Figure \ref{fig:lc}. Apart from PKS 1510-089 and 3C 279, and possibly 4C +21.35 and TON 599, all other sources have only been observed occasionally with (reasonably) dense sampling only around brightening, if any. Thus, statistical estimates may not represent their true essence. Nonetheless, when focusing on VHE emission from FSRQs, which has been reported mainly during brighter MeV-GeV and optical phases, 
observation during a similar brighter or a different phase without a concurrent VHE activity will offer/allow a comparative study which is the central focus of our work.

\subsection{Temporal variability studies}
\label{subsec:tVar}

Variability is the primary defining characteristic of blazars and AGNs and has been the main approach to exploring highly compact sources. Apart from the Bayesian block which marks brightness changes and can be used as a proxy for activity provided the source is well-sampled within a given time frame, a more quantitative statistical estimate is the intrinsic variance and the brightness amplitude  (maximum to minimum ratio) variation. The maximum and minimum flux of each source and their ratio are tabulated in Table \ref{table:ratio} which shows 2-10 times flux variation except for 3C 279, which changed drastically by 24 times the minimum flux. For intrinsic variation, we employed the fractional variability method \citep{2003MNRAS.345.1271V} that measures the variation with respect to the mean and thus, is more meaningful for sources of different brightness levels. It is defined as:
\begin{equation}
    F_{var} = \sqrt{\frac{S^2 - \bar{\sigma_F}^2}{\bar{F}^2}}
\end{equation}
Here $\rm S$ is the variance, $\bar{F}$ is the mean flux, and $\bar{\sigma_F}$ is the mean error in the observed flux. The $F_{var}$ calculated using the above equation is also listed in Table \ref{table:ratio}. The $F_{var}$ shows significant temporal variability. The most variability for 3C 279 with $F_{var}$ 0.60 and the least being S3 0218 with $F_{var}$ 0.12. The latter has too few observations and so may not be reliable, and rather 4C +21.35 has the lowest $F_{var}$
of 0.19. The rest seems roughly similar except PKS 1441+25 which is high due to observation only during the reported VHE flare.

\begin{table}[hbt!]
\caption{The nine VHE FSRQs with their minimum and maximum flux and the ratio (unit of flux - erg $cm^{-2}$ $s^{-1}$) in the first three columns, last column represents the fractional variability amplitude (see section \ref{subsec:tVar})}
\centering
    \begin{tabularx}{\textwidth}{CCCCC} 
     \toprule
     \textbf{Source name} &  \textbf{Min Flux} &
\textbf{Max Flux} & \textbf{Ratio} & \textbf{$F_{var}$}
 \\  
     \midrule
     PKS 0736+017 &  3.27e-12 & 8.48e-12 & 2.6 & 0.27 \\ 
     PKS 1510-089 &  3.76e-12 & 1.81e-11 & 4.8 & 0.25 \\ 
      4C +21.35 &  1.68e-12 & 7.04e-12 & 4.2 & 0.19\\ 
     3C 279 &  4.67e-12 & 1.13e-10 & 24.2 & 0.60\\ 
     B2 1420+32  &  1.51e-12 & 6.48e-12 & 4.3 & 0.27\\ 
     TON 0599 &  7.63e-13 & 7.58e-12 & 9.9 & 0.25\\ 
     PKS 1441+25 &  5.63e-13 & 3.51e-12 & 6.2 & 0.56\\ 
     S3 0218+35 &  8.663-13 & 3.08e-12 & 3.5 & 0.12\\ 
     PKS 0346-27 &  9.14e-13 & 6.00e-12 & 6.6 & 0.30\\
    \bottomrule
    \end{tabularx}
\label{table:ratio}
\end{table}

\begin{figure}
\begin{adjustwidth}{-\extralength}{0cm}
\centering\includegraphics[width=15.5cm, height=9cm]{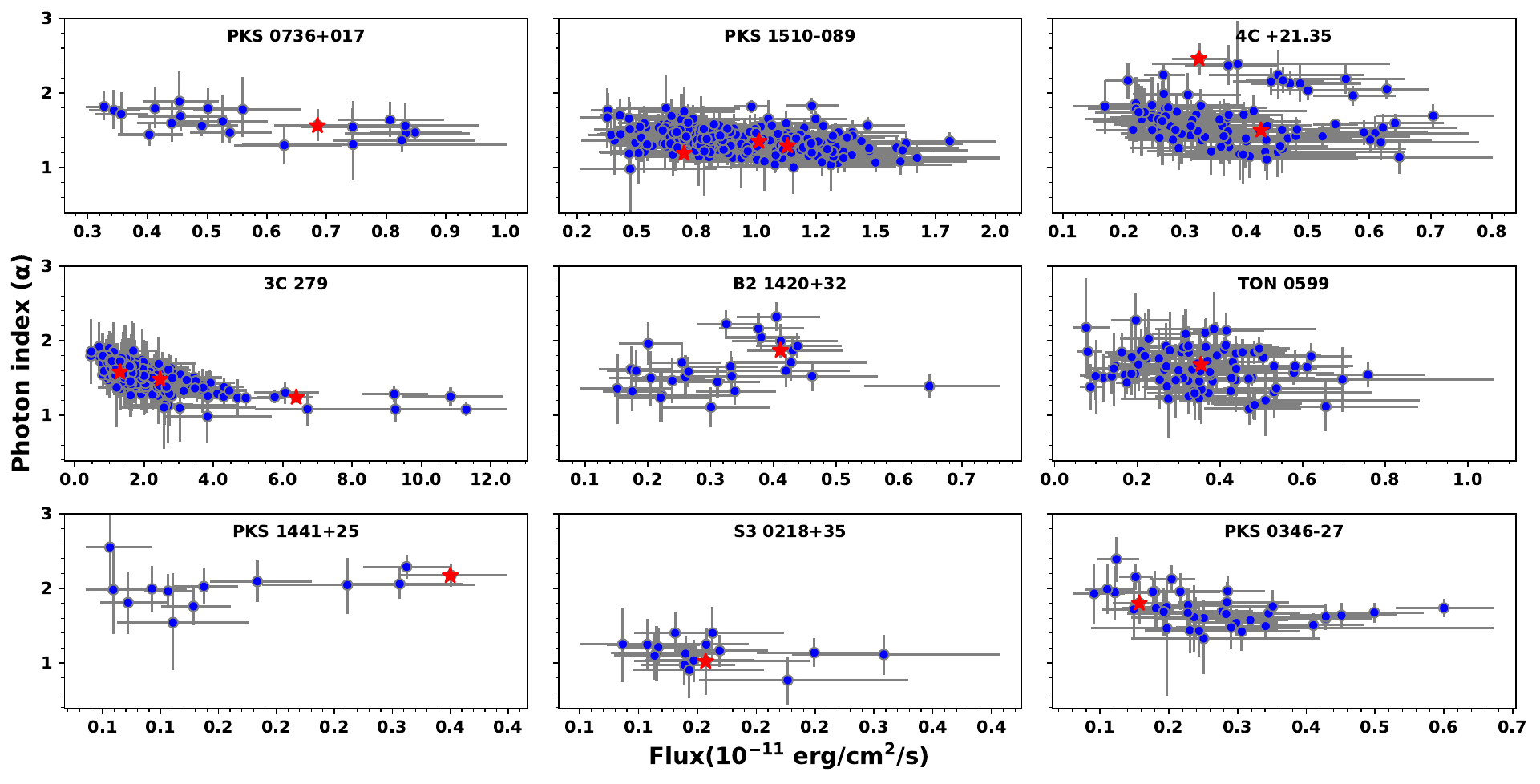}
\caption{Spectral index vs flux of all the nine VHE FSRQs in increasing order of redshift from left to right. The red stars mark the X-ray data associated with the reported VHE activity.}
\label{fig:fsi1}
\end{adjustwidth}
\end{figure}

\subsection{Spectral variability studies}

Brightnening of sources is directly related to a competition between injection of energetic particles, acceleration of particles, and radiative losses. Thus, analysis of the photon spectral index with time and flux state provides insight into the timescales involved and the relative importance of the gains and losses. The right panel of Figure \ref{fig:lc} shows the photon spectral index as a function of time. The photon spectral index varies between 0.5 to 3. We have also plotted the photon spectral index versus photon flux to study the relationship between them in Figure \ref{fig:fsi1}. The photon spectral index at the time of VHE emission is marked using a red star. Except for B2 1420+32, PKS 1441+25, and S3 0218+35, the rest of the VHE FSRQs show/indicate an anti-correlation between flux and spectral index i.e. bluer-when-brighter trend. B2 1420+32 data indicate an anti-correlation but the limited data suggest two different tracks depending on the flux level. A similar behaviour is apparent in 4C +21.35 also. As stated earlier, if one focuses on flux only vis-a-vis VHE activity, it is clearly apparent that there is no correlation between the X-ray flux state and VHE emission and similarly neither with the X-ray spectral state. For sources with multiple VHE detections, the plot instead indicates that the spectral index spans almost the entire range exhibited by the source.

\subsection{Histogram}

\begin{figure}
\begin{adjustwidth}{-\extralength}{0cm}
\centering\includegraphics[width=15.5cm, height=9cm]{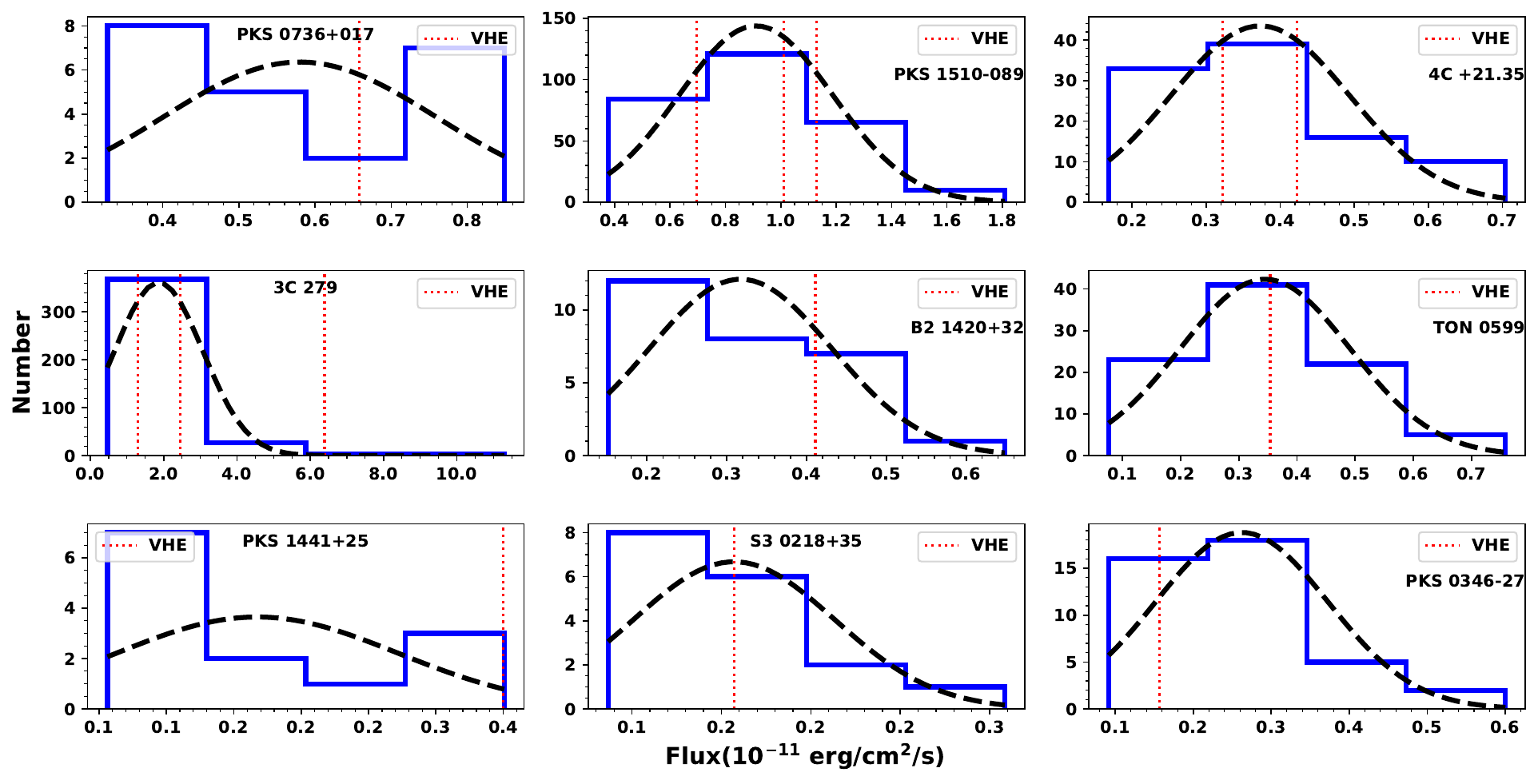}
\caption{Histogram of the flux. The histogram is fitted with the normal distribution shown in the black dashed line. The vertical red dotted line represents the X-ray flux (0.3-10 keV) associated with the VHE activity.}
\label{fig:hf}
\end{adjustwidth}
\end{figure}

\begin{figure}
\begin{adjustwidth}{-\extralength}{0cm}
\centering\includegraphics[width=15.5cm,height=9cm]{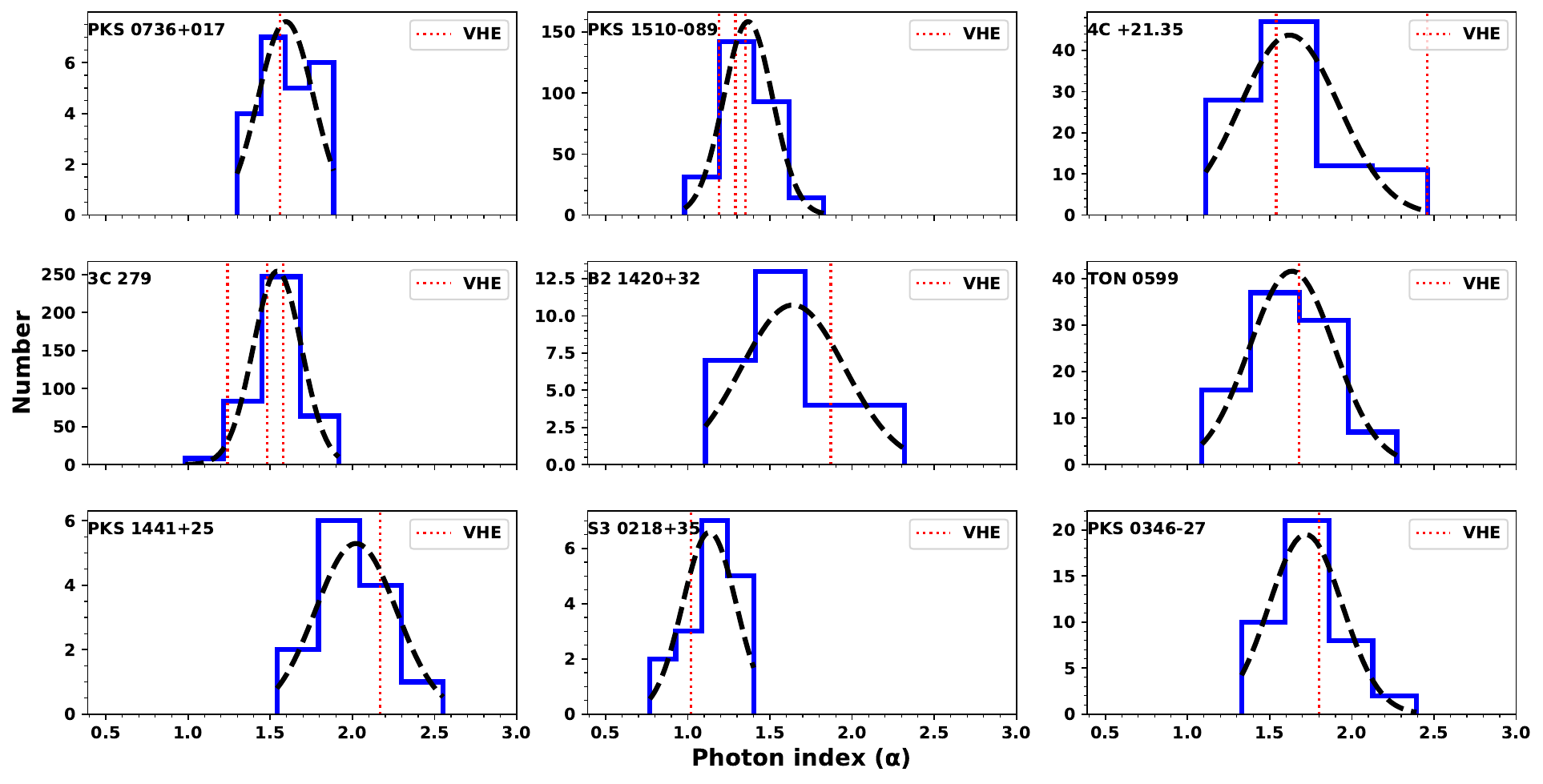}
\caption{Histogram of the spectral energy index. The histogram is fitted with the normal distribution shown in the black dashed line. The vertical red dotted line represents the X-ray spectral index associated with the VHE  activity.}
\label{fig:hsi}
\end{adjustwidth}
\end{figure}

Histograms provide a means to infer the most preferred state for the considered duration in addition to providing insight into variability and range of the variations, if any. The flux and spectral histograms (normalized) are shown respectively in Figure \ref{fig:hf} and Figure \ref{fig:hsi}, and the corresponding normal fit is also plotted. The time of VHE detection is marked with the red dotted line. The Knuth binning method \citep{2019DSP....9502581K} was used to get the optimal number of bins. Within the available information, three interesting facts can be seen. First, 3C 279 -- the most active and variable source of all, the VHE activity is seen only during low X-ray flux states and its flux histogram has fewer bins. Second, contrary to the rest of the sources, PKS 1441+25 has in general much softer X-ray spectra of all and even during the VHE, and third, out of all S3 0218+35 X-ray spectra is the hardest.

\subsection{Spectral Energy Distribution studies}

\begin{figure}
\begin{adjustwidth}{-\extralength}{0cm}
\centering\includegraphics[width=15.5cm,height=9cm]{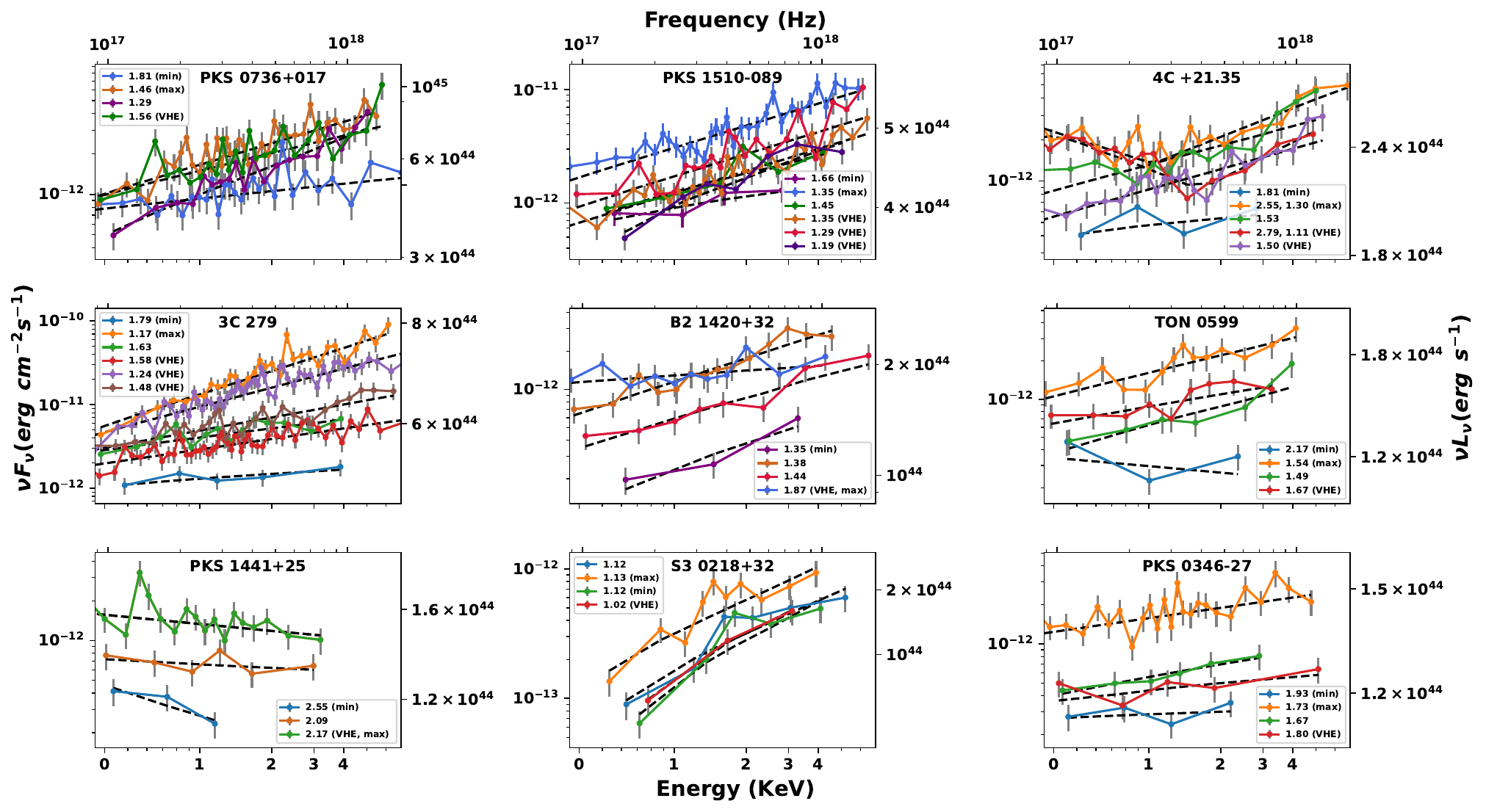}
\caption{A view of different X-ray SEDs exhibited by the VHE FSRQs including the that of minimum (min) and maximum (max) X-ray flux states as well as that of the VHE activity. The number in the label is the best-fit power-law photon spectral index (eq. \ref{eq1}).}
\label{fig:sed}
\end{adjustwidth}
\end{figure}

The X-ray SEDs corresponding to the lowest and the highest recorded flux state along with that of the VHE activity are shown in Figure \ref{fig:sed}.
A power-law model modified with Galactic absorption (tbabs*pow) provides a good description of the X-ray spectrum of all the sources
except one observation of 4C +21.35 for which a broken power-law fits better. All the SEDs were corrected for the Galactic nH value by taking the ratio of unabsorbed flux (nH value set to 0) to absorbed flux (nH value set to galactic nH of the source). In the spectrum too, there is no obvious correlation of X-ray spectral state with VHE state and the spectrum covers all the range between the minimum and maximum. Out of all, PKS 1441+25 only has softer spectra of all. The only other such clear softer spectrum case is of the observation of 4C +21.35 described by a broken power-law with a softer lower energy spectrum (before break). On the other side, S3 0218+35 has harder X-ray spectra of all.


\section{Discussion}
\label{dis}

FSRQs exhibit prominent BLR lines and many show prominent IR-torus emission signature \citep[e.g.][]{2007A&A...473..819R,2008ApJ...672..787K,2011ApJ...732..116M} and thus, FSRQ's gamma-ray emission is generally explained by IC-BLR and IC-IR scenario. However, for VHE emission, IC-BLR is irrelevant due to the onset of KN, and also BLR offers strong opacity to the VHE photons via photo-pair process thus, this combined with the size of BLR under standard AGN paradigm ($<$ a parsec) has been used to argue that VHE emission happens at parsec scales \citep[e.g.][]{2016ApJ...821..102B}. In the case of a single emission region and no SED peak shift, the VHE emission through IC-IR corresponds to the high-energy tail of the particle spectrum which is directly related to the optical-UV synchrotron spectrum. The extension of optical-UV synchrotron spectrum to X-rays can lead to the change in X-ray spectrum, thereby making X-ray a potential window to explore and study such connections and thus, indirectly plausible implications to VHE emission
 \citep[e.g.][]{2015ApJ...815L..23A,2017MNRAS.470.2861S}. \\


Our exploration of X-ray variability using the Swift-XRT data shows strong flux variability by a factor of 2-24 between the minimum and the maximum (see Fig. \ref{fig:lc}), as also indicated by the fractional variability which is between 0.12 to 0.60. However, given the sampling and tendency of the {\it Swift} facility in the autonomous mode (the general operational strategy) to follow transient (brightening) events, the reported flux amplitude variation thus represents a lower limit only. Together with the reported VHE episodes, we find that there is no obvious/straightforward correlation of VHE activity either with the X-ray flux (high/low) or with the spectral index. Instead, VHE emission is somewhere between maximum and minimum X-ray flux, except for PKS 144+25, where the time VHE emission happened coincided with its highest flux. \\

The flux versus spectral energy index shows the "bluer-when-brighter" (anti-correlation i.e., the spectral energy index decreases with higher flux) trend commonly observed in FSRQs \citep{2011JApA...32...87G} for all sources except B2 1420+32, PKS 1441+25, and S3 0218+35. However, if we focus on the VHE emissions, there is no clear correlation between the X-ray spectral states and VHE emissions, though it is generally associated with the harder spectral index. In a few cases, the X-ray spectrum is softer during VHE emission, like 4C+ 21.35 (Refer Figure \ref{fig:fsi1}) and the FSRQ PKS 1441+25 for which all the existing observations correspond to a softer X-ray spectrum. \\

The histograms show a continuous distribution of flux and photon spectral indices of all sources (Refer Figure \ref{fig:hf} and \ref{fig:hsi}). The time of VHE emission does not show any particular trend in the flux histograms. In terms of spectral index, S3 0218+35 is different from the rest with relatively harder spectra. Again, the time of VHE detection does not show any peculiarity. \\

The non-thermal part of optical-UV SEDs\footnote{FSRQs additionally exhibit a prominent blue bump in general which is more apparent during low optical
brightness states e.g. \citep{2007A&A...473..819R,2008ApJ...672..787K}} of the FSRQs which is synchrotron emission generally follows a declining power-law (or more complex e.g.
log-parabola) profile. X-ray lies at the farther end of this spectrum and thus, if optical-UV synchrotron continues unhindered to X-ray energies, it can make
the X-ray spectrum softer depending on the level of emission (e.g. 4C +21.35 broken power-law vis-a-vis other; PKS 1510-089 \citep{2023ApJ...952L..38A}, \citep{2024arXiv241005783K}). As
clear from Figure \ref{fig:sed}, majority have
a rising X-ray SED indicating non-synchrotron origin or negligible synchrotron contribution. In the leptonic scenario, the rising X-ray SED is generally
explained by SSC (e.g. \citep{2023ApJ...952L..38A, 1998MNRAS.301..451G,2016A&A...595A..98A,2016MNRAS.458...56W,2017SSRv..207....5R}). However, it is
interesting to note that many of these have relatively softer/flat spectrum, especially during low X-ray flux states (4C +21.35, 3C 279, TON 0599, PKS 0346-27),
indicating possible synchrotron contribution. PKS 1441+25 is unique among the VHE FSRQs with all the available observations\footnote{very limited X-ray observation, only
around the reported VHE activity} showing a softer X-ray spectrum, attributed to synchrotron
as the dominant mechanism \citep{2015ApJ...815L..23A}, while one of the observations of 4C +21.35 is better explained
by a broken power law with a softer low-energy part ($<2$ keV) and harder (rising) above it, indicating optical-UV synchrotron emission continuing
to X-ray energies \citep{2016MNRAS.458...56W}. 
Such continuations of optical-UV synchrotron into X-ray have direct implications on gamma-ray emission, especially VHE (e.g. \citep{2023ApJ...952L..38A,2015ApJ...815L..23A,2017MNRAS.470.2861S}) and implies a continuation of IC spectrum by the same factor to higher energies
in the Thomson scattering limit in the leptonic scenario -- currently the favored scenario observationally. 
For instance, the explanation of PKS 1510-089 VHE emission during low and high states requires an optical-UV synchrotron spectrum extending well into X-ray regime \citep{2023ApJ...952L..38A}. Apart from PKS 1441+25,  the other odd/peculiar source is S3 0218+32 with a harder X-ray spectra of all. Multi-wavelength modeling in \citet{2016A&A...595A..98A} explains it as an additional HBL-like component but with Compton dominance \citep{2013ApJ...763..134F} like FSRQs, i.e. IC component peak $\geq$ 1 order of magnitude compared to the corresponding synchrotron component while HBL has Compton dominance $\leq$ 1. \\

VHE/TeV sources, regardless of their spectral class, have been explored at X-rays using {\it Swift} data by \citet{2016MNRAS.458...56W}. However, the
study focused only on sources with high signal-to-noise ratios (SNRs) and used averaged spectra to determine and compare different phenomenological spectral models. However, it should be noted that averaging may introduce bias in spectral parameters if the observations to be averaged/combined have significantly
different spectral and brightness states. Focusing on FSRQs, the above-stated study had only five FSRQs among which only three (4C +21.35, 3C 279, PKS 1510-089) had high SNR and thus, were explored spectrally, with 4C +21.35 broken power-law X-ray spectrum already reported and argued to be synchrotron contribution leading to a softer spectrum at lower energies ($\sim$ 1 keV) and harder above it (upturn in SED). In the current work, the VHE FSRQs have increased to nine\footnote{the 10th one was reported recently} and we have focused on general spectral and temporal behavior. In addition to 4C +21.35, we found four more sources: 3C 279, TON 0599, PKS 1441+25, \and PKS 0346-27, with a relatively softer or flat spectrum indicating possible synchrotron contribution, specifically, during low X-ray flux states in 4C +21.35, 3C 279, TON 0599 and PKS 0346-27 while all the observed flux states in PKS 1441+25 is entirely attributed to synchrotron emission (e.g. \citep{2015ApJ...815L..23A,2017MNRAS.470.2861S}). If sources with softer X-ray spectrum, indicating synchrotron contribution correspond to an underlying low-state emission, these sources could be potential persistent VHE sources for the upcoming and future improved sensitivity VHE facilities e.g. the Cherenkov
Telescope Array (CTA, \citep{2013APh....43....3A}) and potential sources for exploration of other research topics (e.g. \citep{2024MNRAS.tmp.2254A,2024Galax..12...34Y}). Though being at high redshift, the steepening due to EBL is expected to be higher.

\section{Conclusion}
\label{con}

We carried out temporal and spectral analysis of the nine VHE FSRQs at X-ray using the {\it Swift}-XRT data with a focus on the comparative study of X-ray behaviour during VHE to those of non-VHE episodes, and to look for markers to identify potential VHE sources. It should be noted that majority of the sources have very limited observations, mostly around VHE events and thus, the data may not be good for general studies but the good sampling around VHE provides sufficient minimal information to explore our focus without much bias, though the availability of more observations will offer better confidence in the inferences. \\

We found strong flux variability -- a factor of 2-24 between minimum and maximum. However, this variability must be considered the least or most conservative given the non-uniform sampling and observational bias towards brighter states by the {\it Swift} facility, thereby possibly missing the observations corresponding to the low X-ray flux states. The temporal activity was further investigated using the Bayesian block and fractional variability amplitude was also estimated. Except, for 3C 279 all the sources seem to have roughly similar fractional variation. Investigation of spectral properties with X-ray brightness shows an anti-correlation between flux and spectral index, widely referred to in the literature as bluer-when-brighter trend, in all with reasonably good sampling. \\

We do not find any correlation between VHE activity either with the X-ray flux or the spectral state of the sources. Interestingly, however, we find 
many FSRQs with a relativity flat or softer X-ray spectrum during low X-ray flux states (4C +21.35, 3C 279, TON 0599, PKS 1441+25, and PKS 0346-27) indicating a continuation of optical-UV synchrotron spectrum into X-rays. If this extended component corresponds to an underlying low-level emission or represents an overshadowed component, then these sources could be potential persistent VHE candidates for future observatories
like CTA. Further, of all FSRQs, PKS 1441+25 and  S3 0218+3 seem interesting and unique with spectral properties quite different from the rest. The former with observations available to date has only softer X-ray spectrum while the latter has the hardest X-ray spectra of all. However, it should be noted that PKS 1441+25 observation is mostly around its reported VHE activity while the latter too is observed occasionally.




\vspace{6pt} 




\authorcontributions{LD performed all the required tasks -- data reduction and analysis, results, and wrote related part of the manuscript under the
supervision of PK. All authors have read and agreed to the published version of the manuscript.}

\funding{This research received no external funding.}

\dataavailability{The {\it Swift}-XRT X-ray data and software needed for reduction are publicly available through HEASARC: \url{https://heasarc.gsfc.nasa.gov/} (Archive and Software).} 

\acknowledgments{P.K. acknowledges support from the Department of Science and Technology (DST), the Government of India, through the DST-INSPIRE faculty grant (DST/ INSPIRE/04/2020/002586).}

\conflictsofinterest{The authors declare no conflicts of interest.} 


\begin{adjustwidth}{-\extralength}{0cm}

\reftitle{References}


\bibliography{vheFsrqX_sorted}

\PublishersNote{}
\end{adjustwidth}

\end{document}